\documentclass[pra,aps,twocolumn,showpacs,superscriptaddress]{revtex4-1}

\usepackage{epsfig}
\usepackage{mathrsfs}
\usepackage{amsmath}
\usepackage{float}
\usepackage{color}
\usepackage{epstopdf}

\begin{document}

\title{Past of a quantum particle and weak measurement}

\author{Zheng-Hong Li}

\affiliation{
	Institute for Quantum Science and Engineering (IQSE)
and Department of Physics and Astronomy, Texas A\&M University, College Station, Texas 77843-4242
}
\affiliation{The National Center for Mathematics and Physics, KACST, P.O.Box 6086, Riyadh 11442, Saudi Arabia}
\affiliation{Beijing Computational Science Research Center, Beijing 100084, China}

\author{M. \surname{Al-Amri}}
\affiliation{
	Institute for Quantum Science and Engineering (IQSE)
and Department of Physics and Astronomy, Texas A\&M University, College Station, Texas 77843-4242
}
\affiliation{The National Center for Mathematics and Physics, KACST, P.O.Box 6086, Riyadh 11442, Saudi Arabia}

\author{M. \surname{Suhail Zubairy}}

\affiliation{
	Institute for Quantum Science and Engineering (IQSE)
and Department of Physics and Astronomy, Texas A\&M University, College Station, Texas 77843-4242
}

\affiliation{Beijing Computational Science Research Center, Beijing 100084, China}

\email{}

\date{\today}

\begin{abstract}

We present an analysis of a nested Mach-Zehnder interferometer in which an ensemble of identical pre- and postselected particles leave a weak trace. A knowledge of the weak value partially destroys the quantum interference. The results, contrary to some recent claims, are in accordance with the usual quantum mechanical expectations.

\end{abstract}

\maketitle

\newcommand{\ds}{\displaystyle}
\newcommand{\dd}{\partial}
\newcommand{\be}{\begin{equation}}
\newcommand{\ee}{\end{equation}}
\newcommand{\beq}{\begin{eqnarray}}
\newcommand{\eeq}{\end{eqnarray}}
\newcommand{\dt}{\ds\frac{\dd}{\dd t}}
\newcommand{\dz}{\ds\frac{\dd}{\dd z}}
\newcommand{\D}{\ds\left(\frac{\dd}{\dd t} + c \frac{\dd}{\dd z}\right)}

\newcommand{\w}{\omega}
\newcommand{\W}{\Omega}
\newcommand{\g}{\gamma}
\newcommand{\G}{\Gamma}
\newcommand{\E}{\hat E}
\newcommand{\s}{\sigma}
\newcommand{\bra}{\langle}
\newcommand{\ket}{\rangle}


\section{Introduction}
Weak measurement \cite{weak1,weak2}, as its name implies, is a kind of quantum measurement where the coupling between the measured system and the measuring device is so weak that the system  remains  unaffected during the process of measurement. A single measurement does not provide any information about the system but, after a large number of repeated measurements on an ensemble of identically prepared pre- and postselected systems, information can be extracted. The notions of weak measurement and weak values were first introduced in a classic paper by Aharonov,  Albert and Vaidman in 1988 \cite{weak1}. Since then this idea has found a number of interesting applications in quantum measurement \cite{Ritchie,Wiseman,Howell,Molmer,science}.

The weak measurements and the physical meaning of weak values remain a subject of arguments and discussion \cite{Leggett,Peres,reply,Hardy1,Hardy2,box1,box2,PRA}. An example is shown in Fig. 1 where a small Mach-Zehnder interferometer is inserted into one arm of a big interferometer. Let us assume that, if a single photon is sent into the small interferometer at position F, the detector $D_3$ clicks with unit probability as a result of interference. In this case, if a photon is sent at the input  and the detector $D_1$ clicks, it is reasonable to assume that the single photon must have followed the outer path A and the probability of its existence inside the smaller Mach-Zehnder interferometer (along paths F, B, C, and E) must be zero. This observation lies at the recent schemes for counterfactual computation \cite{nature} and communication \cite{ours}. However, it is argued in some recent papers \cite{V1,V2} that this conclusion may not be correct. Using the concept of weak measurement it is shown that, in the case the detector $D_1$ clicks, the probability of finding the photon is zero at locations F and E but is non-zero inside the small interferometer along paths B and C. In the words of Vaidman\cite{V1}, ``The photon did not enter the interferometer, the photon never left the interferometer, but it was there''. This is, to put it mildly, very surprising and this surprise is compounded by the claim that the photon number in one arm of the small interferometer is 1 whereas in the other arm its value is -1. As a result, the probability of finding the photon at position E is zero.

The objective of this paper is to resolve this mystery. We show that, although the method of weak measurement provides a very different angle to view quantum systems, both its mathematics and internal essence obey the usual approach to quantum mechanics. For the same quantum system, without the post-selected state, quantum mechanics provides us the probabilities associated with observable quantities. We show that the same is true even when a weak measurement is made in a post-selected system.

In the following, we will first present some general arguments to understand the consequences of the weak measurement on the system. We will then consider an example of system-meter interaction that can possibly be implemented via a dispersive atom-field interaction. Our analysis shows that the disturbance caused by the ``weak'' measurement partially destroys the interference and the probability of finding photon at E is not zero anymore. This disturbance caused by a weak measurement is non-zero no matter how weak the measurement is. We also show that all the results can be understood within the framework of a conventional quantum mechanical approach and the weak measurement does not add anything further in our understanding and interpretation of the system considered in Fig. 1.

\section{Weak Measurement in a Double Mach-Zehnder Interferometer}

First we present a brief discussion about the weak measurement. Suppose there is a pointer $\text{P}$ that  is coupled to an observable $\text{A}$ of a system via a Hamiltonian$H=\hbar \eta \text{AP}$ with a very weak coupling $\eta $. The Hamiltonian perturbs the system state before measurement (pre-selected state $\left| {{\psi }_{i}} \right\rangle $) as ${{e}^{-iH\tau/\hbar}}\left| {{\psi }_{i}} \right\rangle \approx \left| {{\psi }_{i}} \right\rangle -( iH\tau/\hbar) \left| {{\psi }_{i}} \right\rangle =\left| {{\psi }_{i}} \right\rangle -i\eta t\text{A}\left| {{\psi }_{i}} \right\rangle \text{P}$.  To extract the perturbation, we can project it on a post-selected state $\left| {{\psi }_{f}} \right\rangle $, which is independent of  the measurement and allows us to investigate the system with the special final state. This can also be understand as a precondition. As a result, the weak value is defined by $A_w=\left\langle  {{\psi }_{f}} \right|\text{A}\left| {{\psi }_{i}} \right\rangle /\left\langle  {{\psi }_{f}} | {{\psi }_{i}} \right\rangle $. We note that the weak value is not a directly observable quantity in any real experiment and is to be inferred from the data of an actual experiment.

To show the concept more clearly, we  again turn to the example as shown in Fig. 1. We consider the state of the photon at four stages as shown by the dotted lines. A single photon is sent into the left side of the beam-splitter $BS_1$ whose function can be described as
\begin{equation}
    \label{f2}
    U_{L1} :
    \begin{cases}
    \left| 100 \right\rangle \to r \left| 100 \right\rangle + t \left| 010 \right\rangle \\
    \left| 010 \right\rangle \to r \left| 010 \right\rangle - t \left| 100 \right\rangle
    \end{cases}
\end{equation}
where $r$ and $t$ are the reflectivity and transmissivity of $BS_1$ and the photon number state $\left| {{n}_{1}},{{n}_{2}},{{n}_{3}} \right\rangle $ describes the number of photons in the path at the left side of $BS_1$, the path between two $BS_1$s and two $BS_2$s (including position F, B, E) and the path at the right side of $BS_2$. For $BS_2$s, the transformation property  is the same as $BS_1$'s except that their reflectivity and transmissivity are equal to $1/{\sqrt2}$.

\begin{figure}[t]
    \begin{center}
    \includegraphics[height=8.50cm,width=0.95\columnwidth]{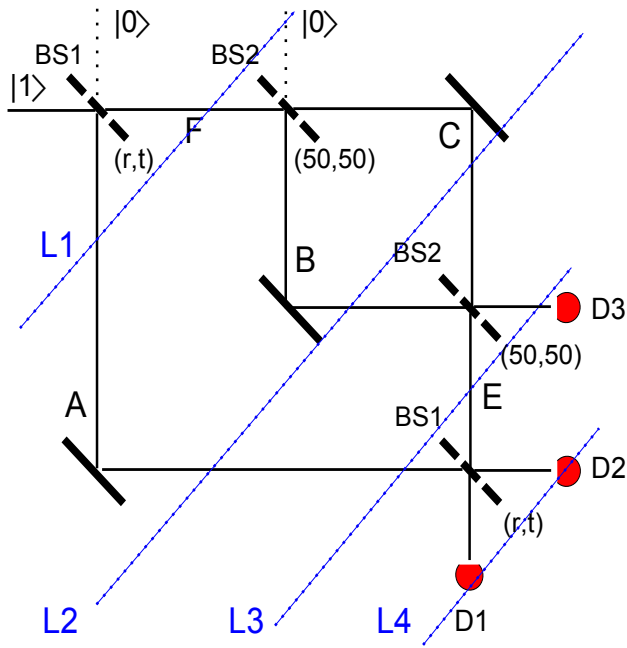}
    \end{center}
   \caption[2Dpattern]
   { \label{fig:fig1}
 (Color online) A small Mach-Zehnder interferometer is added in the upper arm of the big interferometer. A single photon pulse is sent into the setup. $BS$ stands for beam-splitter. $BS_1$'s reflectivity is $r$, transmissivity is $t$ while $BS_2$ is 50\%-50\% beam-splitter. $D_1$, $D_2$ and $D_3$ stand for detectors. L1, L2, L3 and L4 stand for stages corresponding to time evolution of the photon.  }

\end{figure}

At the first stage (L1), the photon state is ${{U}_{L1}}\left| 100 \right\rangle =r\left| 100 \right\rangle +t\left| 010 \right\rangle $. Let the operators ${{U}_{L2}}$, ${{U}_{L3}}$, and ${{U}_{L4}}$ describe operations between the two adjacent stages. We also assume that a weak measurement of the operator $\left| 001 \right\rangle \left\langle  001 \right|$ is made at position C. The pre-selected state at stage L2 is
\begin{equation}
	\left| {{\psi }_{i}} \right\rangle ={{U}_{L2}}{{U}_{L1}}\left| 100 \right\rangle =r\left| 100 \right\rangle +\frac{t}{\sqrt{2}}(\left| 010 \right\rangle +\left| 001 \right\rangle ),
\end{equation}
and, under the condition that $D_1$ clicks, the post-selected state is
\begin{equation}
	\left\langle  {{\psi }_{f}} \right|=\left\langle  100 \right|U_{L4}^{{}}U_{L3}^{{}}=r\left\langle  100 \right|-\frac{t}{\sqrt{2}}   (\left\langle  010 \right|-\left\langle  001 \right|).
\end{equation}

The weak value corresponding to a weak measurement at position C is
\begin{equation}	
{{A}_{C}}=\frac{\left\langle  {{\psi }_{f}} \right|\left| 001 \right\rangle \left\langle  001 \right|\left| {{\psi }_{i}} \right\rangle }{\left\langle  {{\psi }_{f}} \right|\left| {{\psi }_{i}} \right\rangle }=\frac{{{t}^{2}}}{2{{r}^{2}}}
\end{equation}
Similarly, we obtain the weak value corresponding to the weak measurement at position B, which is ${{A}_{B}}=-{{t}^{2}}/(2{{r}^{2}})$. A similar calculation also yields, for the measurements at positions A, E, and F, the weak values ${{A}_{E}}= {{A}_{F}}=0$ and ${{A}_{A}}=1$. These results appear strange at first glance as it seems that the photon appears at B and C but not at E.

Before further discussion, we answer the important question: What is the meaning of the weak value? We note that the denominator in the expression for weak value is $\left\langle  {{\psi }_{f}} \right| \left|{{\psi }_{i}} \right\rangle =\left\langle  100 \right|{{U}_{L4}}{{U}_{L3}}{{U}_{L2}}{{U}_{L1}}\left| 100 \right\rangle $. It is clear that whether we calculate it as $\left\langle  {{\psi }_{f}} \right|\times \left| {{\psi }_{i}} \right\rangle $ or $\left\langle  100 \right|\times \left( {{U}_{L4}}{{U}_{L3}}{{U}_{L2}}{{U}_{L1}}\left| 100 \right\rangle  \right)$ (i.e.,  we let the system evolve stage by stage until stage L4, and then projects it on $D_1$), we  get the same result. The modulus square of $\left\langle  {{\psi }_{f}} \right|\left| {{\psi }_{i}} \right\rangle $ means the probability $D_1$ clicking, which is equal to ${{r}^{4}}$. This implies that the photon passes a classical path without interference. One may argue that the above discussion is meaningless since the measurement is not included. However, in the following we will show that even if we include the measurement process, the concept of post-selected state is still not necessary and the weak value actually tells us the level that the original system is perturbed.

The numerator in the expression for the weak value is $\left\langle  {{\psi }_{f}} \right|\left| 001 \right\rangle \left\langle  001 \right|\left| {{\psi }_{i}} \right\rangle =\left\langle  100 \right|{{U}_{L4}}{{U}_{L3}}\left| 001 \right\rangle  \left\langle  001 \right|{{U}_{L2}}{{U}_{L1}}\left| 100 \right\rangle $. This expression is no different if we calculate ${{U}_{L4}}{{U}_{L3}}\left| 001 \right\rangle \left\langle  001 \right|{{U}_{L2}}{{U}_{L1}}\left| 100 \right\rangle $ at first, and then project it on $D_1$. Nevertheless, the physical meaning of this quantity is clear: The photon state evolves stage by stage until,  at stage L2, a projection measurement is made so that the photon state collapses to $\left| 001 \right\rangle $. After that, the system evolves again but with a new initial state. The interference is destroyed as a result of the projection measurement. The quantity $|\left\langle  {{\psi }_{f}} \right|\left| 001 \right\rangle \left\langle  001 \right|\left| {{\psi }_{i}} \right\rangle {{|}^{2}}$ represent the probability of $D_1$ clicking under the condition that the photon is found at the position C. In this case a click at $D_1$ corresponds to a different situation in comparison with when we did not try to obtain which-path information. Obviously, since we tried to obtain the which-path information, the interference is lost.

In a weak measurement, the  evolution is given by ${{e}^{-iHt}}\left| {{\psi }_{i}} \right\rangle \approx \left| {{\psi }_{i}} \right\rangle -(iHt)/\hbar \left| {{\psi }_{i}} \right\rangle =\left| {{\psi }_{i}} \right\rangle -i\eta t\left| 001 \right\rangle \left\langle  001 \right|\text{P}\left| {{\psi }_{i}} \right\rangle $, and the measurement does not disturb the original system too much. However, the weak value itself is not ``weak'', which comes from the partial system whose interference is destroyed. In other words, the weak value leads to noise that stems from the measurement. In particular, the weak value can not tell us any information about the photon path without affecting next evolution processes of the photon even if the final state does not change.

Now we can roughly answer why ${{A}_{B}},{{A}_{C}}$ are non-zero but ${{A}_{E}}=0$. The main reason is that these three measurements are not made on the same system at the same time. If we do not make any measurement at B or C, there is no doubt no photon will be found at E. A straightforward calculation without post-selected state leads to the same result. However, if we make a measurement at B or C, the situation is different. Since interference is destroyed, the photon has some probability leaking into E and finally causes $D_1$ to click. The probability of finding the photon at E is not zero anymore.

So far we have qualitatively discussed the effect of weak measurement and shown that the approach of weak measurement should not lead to a paradox that does not exist in the usual quantum approach. In particular we have discussed that the weak value describes the noise corresponding to measurement. In the following, we consider a model to show that the conclusion that the photon exists in paths B and C but not E is not correct and the explanation in terms of  negative photon numbers is not needed.

\section{Weak Non-Demolition Measurement}

We consider a weak quantum non-demolition measurement using a Hamiltonian of the form $H=\hbar \eta {a_{c}}^{\dagger}{a_{c}} \left| b \right\rangle \left\langle  b \right|$, where the system operator is $a_{c}^{\dagger}{a_{c}}=\left| 001 \right\rangle \left\langle  001 \right|$ indicating the measurement is at position C and the state $|b\rangle$ is the state of the meter.

Such a Hamiltonian can, for example, be realized by  a single three-level atom in the cascade configuration in the arm C \cite{Brune,atom}. The upper two-levels $\left| a \right\rangle $ and $\left| b \right\rangle $ are dispersively coupled to the photon with a detuning $\Delta $ such that  $\eta =\hbar {{g}^{2}}/\Delta $ \cite{Scully} with $g$ being the atom-field coupling coefficient. The atom, acting as a meter,  is initially prepared in a superposition of the middle level $\left| b \right\rangle $ and the lower level $\left| c \right\rangle $, i. e., $\left| {{\psi }_{A}} \right\rangle =(\left| b \right\rangle +\left| c \right\rangle )/\sqrt{2}$.

According to the photon-atom interaction at position C, the projection of the evolved state on the final state $\left| \alpha \right\rangle \left| {{\psi }_{f}} \right\rangle $ (with $D_1$ clicking and the atom found in level $\alpha$ with $\alpha=b,c$) is
\begin{equation}
\begin{split}
  \left\langle  \alpha \right|\left\langle  {{\psi }_{f}} \right| & {{e}^{-iH\tau /\hbar}}\left| {{\psi }_{i}} \right\rangle \left| {{\psi }_{A}} \right\rangle \\
  &\approx \left\langle  \alpha \right|\left\langle  {{\psi }_{f}} \right|(1-iH\tau/\hbar )\left| {{\psi }_{i}} \right\rangle \left| {{\psi }_{A}} \right\rangle  \\
 & =\left\langle  \alpha \right|\left\langle  {{\psi }_{f}} \right|\left| {{\psi }_{i}} \right\rangle \left( \frac{\left| b \right\rangle +\left| c \right\rangle }{\sqrt{2}}-i{{A}_{C}}\eta \tau \frac{\left| b \right\rangle }{\sqrt{2}} \right) \\
 & \approx \left\langle  \alpha \right|\left\langle  {{\psi }_{f}} \right|\left| {{\psi }_{i}} \right\rangle ({{e}^{-i\eta {{A}_{C}}\tau }}\left| b \right\rangle +\left| c \right\rangle )/\sqrt{2}
\end{split}
\end{equation}
We assume that the atom undergoes the unitary transformation $\left| b \right\rangle \to (\left| b \right\rangle +i\left| c \right\rangle )/\sqrt{2}$, $\left| c \right\rangle \to (i\left| b \right\rangle +\left| c \right\rangle )/\sqrt{2}$ after its interaction with the photon wave packet at C. The amplitude becomes $\left\langle  i \right|\left\langle  {{\psi }_{f}} \right|\left| {{\psi }_{i}} \right\rangle \left[ {{e}^{i\eta {{A}_{C}}\tau }}(\left| b \right\rangle +i\left| c \right\rangle )+(i\left| b \right\rangle +\left| c \right\rangle ) \right]/2$.
The probability of finding the atom in the levels $\left| b \right\rangle $ and $\left| c \right\rangle $ are
\begin{equation}
\label{pbf}
	{{P}_{b}}=\frac{1}{2}|\left\langle  {{\psi }_{f}} \right|\left| {{\psi }_{i}} \right\rangle {{|}^{2}}[1-\sin (\eta {{A}_{C}}\tau )],
\end{equation}
\begin{equation}
\label{pcf}
	{{P}_{c}}=\frac{1}{2}|\left\langle  {{\psi }_{f}} \right|\left| {{\psi }_{i}} \right\rangle {{|}^{2}}[1+\sin (\eta {{A}_{C}}\tau )].
\end{equation}
The weak value $A_C$ can now be inferred from $P_b$ and $P_c$ as follows:
\begin{equation}
	{{A}_{C}}=\frac{1}{\eta \tau }\arcsin \frac{{{P}_{c}}-{{P}_{b}}}{{{P}_{b}}+{{P}_{c}}}.
\end{equation}
We also note that
\begin{equation}
|\left\langle  {{\psi }_{f}} \right|\left| {{\psi }_{i}} \right\rangle {{|}^{2}}={{P}_{b}}+{{P}_{c}},
\end{equation}
Thus the weak value $A_C$ is obtained indirectly under the first order approximation. The probability of $D_1$ clicking does not change as a result of weak measurement since $|\left\langle  {{\psi }_{f}} \right|\left| {{\psi }_{i}} \right\rangle {{|}^{2}}={{P}_{b}}+{{P}_{c}}$. This makes it tempting to claim that the weak measurement has no influence on the final outcome and we should be able to conclude that the photon exists in path C but not E. The situation is however more complex and we look at it  more carefully.

 The evolved state at stage L2 after the measurement is
\begin{equation}	
\begin{split}
  & (1-iH\tau/\hbar )\left| {{\psi }_{i}} \right\rangle \left| {{\psi }_{A}} \right\rangle \\
  & =(1+i)\left[ r\left| 100 \right\rangle +\frac{t}{\sqrt{2}}(\left| 010 \right\rangle +\left| 001 \right\rangle ) \right](\left| b \right\rangle +\left| c \right\rangle )/2 \\
 & -i \frac{t}{\sqrt{2}} \eta \tau \left| 001 \right\rangle \left( \left| b \right\rangle +i\left| c \right\rangle  \right)/2
\end{split}
\end{equation}
Here the last term is the noise term. It is true that, due to the weak nature of interaction ($\eta \tau <<1$), the direct measurement of the atomic levels $\left| b \right\rangle $ and $\left| c \right\rangle $ can not give us significant information of the photon path. However, the distribution of the photon in different paths has changed corresponding to the measurement. The quantum interference is partially destroyed. For example, if the atom is found in level $\left| b \right\rangle $, the probability of finding the photon at E is ${\eta}^2 {\tau}^2 t^2 /16$. Since the component of the photon state at position E corresponding to $\left| b \right\rangle $ and $\left| c \right\rangle $ have different phases, they have different contribution for the interference happening at $BS_1$. In one case the probability of $D_1$ clicking increases, in the other case it decreases resulting in a null effect. However, in the following, we  show that, if we carry out our calculation exactly (instead of restricting only to the first order in $\eta \tau$), the measurement changes the probability of $D_1$ clicking in the order  $(\eta \tau)^2$.

The photon-atom state at stage L2 is given by
\begin{equation}
\begin{split}
\label{f3}
  & {{e}^{-iH\tau }}\left| {{\psi }_{i}} \right\rangle \left| {{\psi }_{A}} \right\rangle  \\
 & =(r\left| 100 \right\rangle +\frac{t}{\sqrt{2}}(\left| 010 \right\rangle +\left| 001 \right\rangle) )\frac{\left| b \right\rangle +\left| c \right\rangle }{\sqrt{2}} \\
 &+\frac{{{t}_{{}}}}{2}\left( {{e}^{-i\eta \tau }}-1 \right)\left| 001 \right\rangle \left| b \right\rangle  \\
 & \to \frac{1+i}{2}(r\left| 100 \right\rangle +\frac{t}{\sqrt{2}}(\left| 010 \right\rangle +\left| 001 \right\rangle) )(\left| c \right\rangle +\left| b \right\rangle ) \\
 & +\frac{{{t}_{{}}}}{2\sqrt{2}}\left( {{e}^{-i\eta \tau }}-1 \right)\left| 001 \right\rangle (\left| b \right\rangle +i\left| c \right\rangle ). \\
\end{split}
\end{equation}
in the last step, we made the same unitary transformation for $\left| b \right\rangle$ and $\left| c \right\rangle$ as discussed above (namely $\left| b \right\rangle \to (\left| b \right\rangle +i\left| c \right\rangle )/\sqrt{2}$, $\left| c \right\rangle \to (i\left| b \right\rangle +\left| c \right\rangle )/\sqrt{2}$). The photon-atom state evolves to
\begin{equation}
\begin{split}
  & {{U}_{L4}}{{U}_{L3}}{{e}^{-iH\tau }}\left| {{\psi }_{i}} \right\rangle \left| {{\psi }_{A}} \right\rangle \\ &={{U}_{L4}}[\frac{1+i}{2}(r\left| 100 \right\rangle +t\left| 001 \right\rangle )(\left| c \right\rangle +\left| b \right\rangle ) \\
 & +\frac{{{t}_{{}}}}{4}\left( {{e}^{-i\eta \tau }}-1 \right)(\left| 001 \right\rangle -\left| 010 \right\rangle )(\left| b \right\rangle +i\left| c \right\rangle )] \\
 & =\frac{1+i}{2}({{r}^{2}}\left| 100 \right\rangle +rt\left| 010 \right\rangle +t\left| 001 \right\rangle )(\left| c \right\rangle +\left| b \right\rangle ) \\
 & +\frac{{{t}_{{}}}}{4}\left( {{e}^{-i\eta \tau }}-1 \right)(\left| 001 \right\rangle -r\left| 010 \right\rangle +t\left| 100 \right\rangle )(\left| b \right\rangle +i\left| c \right\rangle ).
\end{split}
\end{equation}
Regardless of whether we find the atom in level $|b\rangle$ or $|c\rangle$, the probability of finding the photon at E is $\frac{t^2}{4}[1-\cos (\eta \tau )]$. The probability $D_1$ clicking is now given by $P_b+P_c=r^4+\frac{t^2}{2}[ 1-\cos (\eta \tau ) ]\left( \frac{t^2}{2}-r^2 \right)$, where $P_b=\left| \frac{1+i}{2}r^2+\frac{t^2}{4}\left( e^{-i\eta \tau }-1 \right) \right|^2$ and ${{P}_{c}}={{\left| \frac{1+i}{2}{{r}^{2}}+i\frac{t^2}{4}\left( {{e}^{-i\eta \tau }}-1 \right) \right|}^{2}}$. An interesting observation is that, if we project Eq. \eqref{f3} directly on the post-selected state, we get the same result.

Here we note that the probability of $D_1$ clicking is changed due to the influence of measurements. In the weak measurement approximation, these results reduce to ${{P}_{b}}\approx r_{{}}^{4}[1-\sin ({{A}_{C}}\eta \tau )]/2$  and ${{P}_{C}}\approx r_{{}}^{4}\left[ 1+\sin ({{A}_{C}}\eta \tau ) \right]/2$ yielding the  same results as given in Eqs. \eqref{pbf} and \eqref{pcf}. The second order difference in the clicking rate of $D_1$ which comes from measurement disappears.

So here is the resolution of the confusion. First we note that, although the probability of finding the photon at E is second order in $\eta \tau$, the amplitude is still first order. This linear amplitude proportional to the transmissivity ($-i\eta \tau t^2/4$), when added to the amplitude along path A ($r^2 (1+i)/2)$, yields $P_b$ that is linear in $\eta \tau$. A similar result is obtained for $P_c$. This important observation clarifies that, no matter how weak the quantum non-demolition measurement at B or C is, it destroys the interference leading to a non-zero amplitude at E, which when combined with the amplitude along A gives the correct detection probability at detector $D_1$. As $P_E$ appears to be zero in the linear approximation, we can be  led to the erroneous claim that the photon does not exist at E when a weak measurement is made at B or C. However we see that it is the linear amplitude at E that is responsible for the experimentally observable quantities $P_b$ and $P_c$ in the linear order.

Up to now, we proposed a detailed design of making a weak measurement at location C to show that the essence of the weak measurement is not different from usual quantum mechanics methods. More importantly, we show that, no matter how weak the measurement is, a measurement always disturbs the original system. This is the price we always need to pay in order to get the information in an interferometric system of the type shown in Fig. 1. Thus we obtain contribution not only from the original system (from path A in our case) but also the perturbation coming form measurements we made (in path B and/or C).

In order to emphasize our point, we can consider another measurement on the same system at the same time. For example, we can place the meter atoms at positions C and E and carry out the joint measurement. If Vaidman is correct \cite{V1,V2}, the pointer at E should not find any photon. However, even when we do the calculation that follows the logic of weak measurement, the corresponding joint weak value  yields a different result in accordance with the usual quantum mechanical expectation. The joint weak value \cite{joint} corresponding to a click at $D_1$ is given by
\begin{equation}
\frac{\left\langle  {\psi }_{f}  \right|{{U}_{L4}}\left| 010 \right\rangle \left\langle  010 \right|{{U}_{L3}}\left| 001 \right\rangle \left\langle  001 \right|{{U}_{L2}}{{U}_{L1}}\left| \psi_i  \right\rangle }{\left\langle  {\psi }_{f}  \right|{{U}_{L4}}{{U}_{L3}}{{U}_{L2}}{{U}_{L1}}\left| \psi_i  \right\rangle }=\frac{{{t}^{2}}}{2{{r}^{2}}}
\end{equation}
This value is non-zero and is equal to ${{A}_{C}}$. The reason is that if the photon is found at C (and finally causes $D_1$ to click), it must pass through path E.

\section{Concluding Remarks}

In summary, we have considered a generic system of measurement of weak values in a nested Mach-Zehnder interferometer and have shown, via a straightforward quantum mechanical calculation, that the weak measurement and the subsequent evolution are consistent with our quantum mechanical expectations. There is no mystery or paradox in the simple set up of Fig. 1. The quantum mechanical paradigm that a measurement disturbs the system can explain the outcome of a potential experiment. We note that we considered a particular model for the system-meter interaction to illustrate our results but similar conclusions can be drawn in other systems such as in \cite{box2}. We also mention that the three-box paradox \cite{weak2} corresponds to the special case $r=1/\sqrt{3}$ and $t=\sqrt{2/3}$. A resolution of this paradox can be analyzed within the framework presented in this paper. For a careful analysis of weak measurement, the effect of the measuring device or the meter should be included in the subsequent evolution.

\begin{acknowledgments}
This research is supported by an NPRP grant (4-520-1-083) from the Qatar National Research Fund (QNRF) and a grant from King Abdulaziz City for Science and Technology (KACST).

\end{acknowledgments}




\end{document}